\def\ni{\noindent}
\def\Im{I_{{\rm m}}}
\def\xim{\xi_{{\rm m}}}
\def\etam{\eta_{{\rm m}}}
\def\rp{r^\prime}
\def\thetap{\theta^\prime}
\def\scrF{{\cal F}}

\documentstyle[epsf]{mn}

\title[Cusps without chaos]
{Cusps without chaos}
\author[S.~Sridhar and J.~Touma]
	{S.~Sridhar$^1$\thanks{Email:~sridhar@iucaa.ernet.in} and
	 J.~Touma$^2$\thanks{\mbox{Present Address: University of
 	Texas, McDonald Observatory,} \mbox{RLM15.308
         Austin, TX 78712, USA.} \mbox{Email:~touma@harlan.as.utexas.edu}}\\ 
	$^1$~Inter-University Centre for Astronomy and Astrophysics, 
        Ganeshkhind, Pune 411 007, INDIA\\
	$^2$~Canadian Institute for Theoretical Astrophysics,
	60 St. George Street, Toronto MS5 1A7, CANADA}

\date{IUCAA Preprint 49/96, December 1996}
 
\pagerange{\pageref{firstpage}--\pageref{lastpage}}
\pubyear{1996}
 
\begin{document}
\label{firstpage}
 
\maketitle
\begin{abstract}
We present cuspy, non--axisymmetric, scale--free mass models of 
discs, whose gravitational potentials are of St\"ackel
form in parabolic coordinates.
{\em A black hole may be
added at the centre, without in any way affecting the St\"ackel form};
the dynamics in these potentials is, of course, fully integrable.
The surface density, $\Sigma_{disc}\propto 1/r^{\gamma}$,
where $0 <\gamma < 1$ corresponds to steep cusps for which the central
force diverges. 
Thus cusps, black holes, and non--axisymmetry are not a sure recipe for
chaos, as is generally assumed. 
A new family of orbits, {\em lens} orbits,
emerges to replace the box orbits of models 
of elliptical galaxies that have constant--density cores.
Loop orbits are conspicuous by their absence.
Both lenses and boxlets (the other family of orbits), can be elongated
in the direction of the density distribution, a property that is favourable
for the construction of non--axisymmetric, self--consistent equilibrium
models of elliptical galaxies.
\end{abstract}


\section{INTRODUCTION}

More than two decades ago, it became apparent that elliptical galaxies
are not rotationally supported, oblate objects (Bertola \& Cappacioli~1975,
Illingworth~1977). 
Subsequent dynamical investigations (c.f. Binney~1976,
1978, Schwarzschild~1979, 1982) 
led to the picture of ellipticals as slowly/non--rotating, 
triaxial objects with anisotropic velocity distribution, whose self--consistent
gravitational  potentials supported mostly regular orbits.
The discovery of  fully integrable, triaxial mass models
by Kuzmin (1973) and de~Zeeuw \& Lynden--Bell (1985) 
provided analytic power, and opened up a new era of dynamical
investigations. 
The key insight was the employment of ``St\"ackel
potentials'' in ellipsoidal coordinates; in addition to the energy, 
stellar orbits in St\"ackel potentials respect two extra isolating integrals.
A range of triaxial, self--consistent equilibrium configurations could 
now be constructed in an efficient manner (Statler~1987); 
a property common to
all these models is a constant--density core in which stars
execute near--harmonic oscillations (``box orbits''; see de~Zeeuw~1985
for a thorough account of orbits).

Recent observations (c.f. M{\o}ller, Stiavelli, \& Zeilinger 1995,  
Ferrarese et al 1994, Lauer et al 1995) of elliptical galaxies, however,
do not support the notion of a constant--density core; indeed, 
the density of stars appears to rise toward the centre in a 
power--law cusp. It is also widely believed that these galaxies harbour dark  objects, possibly supermassive black holes, at their centres
(c.f. Kormendy \& Richstone~1995). Gerhard \& Binney~(1985) showed that
central black holes, or steep density cusps will destroy box orbits 
in triaxial configurations, and argued that the boxes will
be replaced by chaotic orbits.   
They also noted that chaotic orbits fill a region of space that is too round for strongly non--axisymmetric, self--consistent equilibria to
be constructed; thus it has been suggested that cuspy galaxies 
with black holes must necessarily be axisymmetric, at least near the
centre (c.f. Merritt~1996 for an account based on recent numerical explorations). 
Although the link between cusps, non--axisymmetry, and chaos is supported by 
numerical results in model potentials, we question its generality.  
We construct {\em separable} mass models of cuspy, non--axisymmetric,
scale--free discs, with  central black holes in this paper.
The potentials of these discs promote a new family of orbits, {\em lens}
orbits, to replace the box orbits of galaxies with constant--density
cores. We hope that our  results  will add fuel to the 
ongoing debate on the intrinsic shapes of elliptical galaxies.

\section{POTENTIAL--SURFACE DENSITY PAIRS}

We  begin with an explicit formula for the potential (in the plane)
of our discs, in the usual $(r, \theta)$
polar coordinates:

\begin{equation}
\Phi= K r^{(1-\gamma)}F(\theta) -\frac{GM}{r},
\label{potential}
\end{equation}

\ni where $K >0$ is a constant, and

\begin{equation}
F(\theta)=\left\{(1+\sin\theta)^{(2-\gamma)}
+ (1-\sin\theta)^{(2-\gamma)}\right\}\,.
\label{funcF}
\end{equation}

\ni Knowledge of the potential in the plane of the disc suffices to 
determine the surface density though the integral (c.f. Binney \&
Tremaine~1987),

\begin{equation}
\Sigma(\rp, \thetap)=\frac{1}{4\pi^2 G}\int\frac{\left(\nabla^2\Phi\right)
\,r\,dr\,d\theta}
{[r^2 + r^{\prime\, 2} -2r\rp\cos(\theta -\thetap)]^{1/2}}\,,
\label{sigphi}
\end{equation}

\ni It is evident, from dimensional considerations, that 
$\Sigma(r, \theta)=S(\theta)/r^{\gamma} + M\delta({\bf r})\,$ is the form 
of the surface density arising from a $\Phi$  given by  equation~(\ref{potential}).    
For $0<\gamma <1$, straightforward manipulations of   
equation~(\ref{sigphi}), using 
Legendre  functions $P_{-\gamma}$, demonstrate that\footnote{The 
proof of the second of these relations requires integration by parts,
and the elimination of $P_{-\gamma}^{\prime}$ by using the standard
differential equation satisfied by $P_{-\gamma}\,$ (c.f. Morse \&
Feshbach~1953).}

\begin{eqnarray}
S(\thetap) &=& \frac{K}{4\pi G\sin(\pi\gamma)}\int_{-\pi}^{\pi}\,d\theta
\left\{(1-\gamma)^2 F(\theta) +\right.\nonumber \\[1ex]
&{}&\quad\quad F^{\prime\prime}(\theta)
\left.\right\} P_{-\gamma}[-\cos(\theta -\thetap)]\,,\\[1em]
&=& \frac{K}{8\pi G\sin(\pi\gamma)}\int_{-\pi}^{\pi}\,d\theta
\left\{\sin^2\theta P_{-\gamma}^{\prime\prime}(-\cos\theta) +\right.\nonumber \\[1ex]
&{}&\quad\quad
\left.(1-\gamma)(2-\gamma)P_{-\gamma}(-\cos\theta)\right\}F(\theta + \thetap)\,,
\nonumber\label{funcS}
\end{eqnarray}

\ni is {\em finite and positive}.
Finiteness is guaranteed by the singularity of $P_{-\gamma}(\mu)$
at $\mu= -1$ being integrable, and positivity simply because $F(\theta)$,
$P_{-\gamma}(\mu)$ and
$P_{-\gamma}^{\prime\prime}(\mu)$ are all positive quantities.
If we neglect the contribution from the black hole, 
$0 <\gamma <1$ for all our models implies that the force on a test particle  
diverges, whereas its circular speed goes to zero for small $r\,$.
Isocontours of surface density, and potential 
for $\gamma$ equal to $0.1$ and $0.5$ are displayed 
in Figure~1. The  contours (Figures~1a and 1c)
are guitar--shaped, with the depression 
near the ordinate increasing with $\gamma$; 
of course, the potential isocontours are much rounder
than the density isocontours. The location of the black hole---the origin---is
indicated by the solid dot. While drawing the potential isocontours
(Figures~1b and 1d), we have excluded the contribution of the
black hole, to better display the potentials of the discs 
themselves. The  major axes of the potential isocontours
are exactly twice as long as the minor axes, a property that is readily 
verified using equation~(\ref{funcF}).

A remarkable property of this family of potentials is that all stellar orbits 
are non chaotic. To make this explicit, we will rewrite it in St\"ackel
form in parabolic coordinates.

\begin{figure}
\epsfxsize=3.3truein\epsfbox[76 200 564 690]{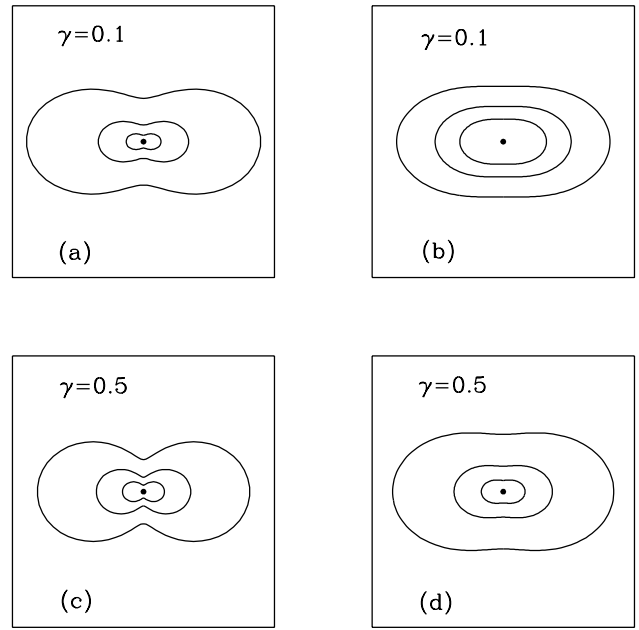}
\caption{ Isocontours of Surface Density and Potential:
Figures (a) and (c) show isocontours of the surface density for
two different values of $\gamma$. The corresponding potentials
are displayed in the panels on the right in (b) and (d).
Successive isocontours in (a) have density ratios of $1.1$, whereas the
ratio is $1.5$ for the other three figures. The location of the
central black hole is shown as a solid dot in all four figures,
although the contribution to the potentials is not included
in (b) and (d). }
\end{figure}

\section{DYNAMICS IN PARABOLIC COORDINATES}
 
Let $\xi = r(\sin\theta + 1)$ and 
$\eta = r(\sin\theta - 1)$ be parabolic coordinates in the plane. Lines of
constant $\xi$ and $\eta$ intersect the $y$--axis at $\xi/2$ and 
$\eta/2$ respectively, and the $x$--axis at $\pm\xi$ and $\pm\eta$ respectively.
Denoting the canonical momenta by $p_\xi$ and $p_\eta$, the  Hamiltonian,

\begin{equation}
H(p_{\xi}, p_{\eta}, \xi, \eta)=
\left(\frac{2\xi}{\xi -\eta}\right)p_{\xi}^2 +
\left(\frac{2\eta}{\eta -\xi}\right)p_{\eta}^2 +\Phi\,,
\label{ham}
\end{equation}

\ni governs dynamics in the $(\xi, \eta)$ plane. Since $\Phi$ 
is independent of time, $H=E$ is a conserved quantity.  
The potential of equation~(\ref{potential}) may be 
written in the St\"ackel form,

\begin{equation} 
\Phi= \frac{\scrF_{+}(\xi)}{\xi - \eta} + \frac{\scrF_{-}(\eta)}{\eta - \xi}\,,\label{stackel}
\end{equation}

\ni To verify this, we choose

\begin{eqnarray}
\scrF_{+}(\xi) &=&2K
\xi^{(2-\gamma)}-GM\,\nonumber \\[1em]
\scrF_{-}(\eta) &=& -2K
|\eta|^{(2-\gamma)}+GM\,, \label{ffstackel}
\end{eqnarray}

\ni and substitute for $(\xi, \eta)$ in terms of $(r, \theta)\,$.
It is well--known that,
for such potentials, the Hamilton--Jacobi
equation separates, yielding an extra conserved quantity
(c.f. Landau \& Lifshitz~1976),

\begin{eqnarray}
I &=&2\xi p_{\xi}^2- E\xi + \scrF_{+}(\xi) \nonumber \\[1em]
&=&2\eta p_{\eta}^2 -E\eta + \scrF_{-}(\eta)\,,\label{third}
\end{eqnarray}

The isolating integrals, $E$ and $I$ may be used
to  classify orbits. For the generic case of a black hole,
plus a cuspy density distribution, $E$ can  assume both
positive and negative values. For fixed $E$, 
the ``second'' integral, $I$, will determine the excursions in $\xi$ and 
$\eta$. The requirement that $p_{\xi}^2$
and $p_{\eta}^2$ be non negative, forces

\begin{equation}
I\geq g(\xi)\,,\quad\quad -I\geq g(|\eta|)\,,
\label{range}
\end{equation}

\ni where

\begin{equation}
g(s)= 2K s^{(2-\gamma)} -Es -GM\,;\quad\quad s\geq 0\,.
\label{funcg}
\end{equation}

\ni For fixed $E$,
the range of $I$ is determined by the minimum of the function $g(s)$,
which is necessarily non--positive. If we denote this minimum value by $-\Im(E)$,
we obtain the condition $-\Im < I < \Im$. Since $s >0$,  

\begin{equation}
\Im(E) = \left\{  \begin{array}{ll}
			    			CE^{\left(\frac{2-\gamma}{1-\gamma}\right)} +GM  & \mbox{if $E \geq 0$} \\[1em]

GM  & \mbox{if $E \leq 0\,$,} \label{funcgm}
\end{array} 
\right.
\end{equation}

\ni where $C=(1-\gamma)[2K(2-\gamma)]^{1/(\gamma -1)}/(2-\gamma)$
is a positive constant.
Let us consider two cases (the illustrative figures for which,
given in Figures~2a--2d, are drawn for $\gamma=0.5$):

\ni (i) $E < 0\:$: We find that $g(s)$ is a monotonically increasing function 
of $s$, attaining a minimum value of $-GM$ for $s=0$ 
(see Figure~2a). This implies
that $-GM < I < GM$. For a fixed (say, positive) value
of $I$, the motion is bounded by the coordinate curves
$0\leq \xi \leq \xim$, and 
$0\leq |\eta|\leq\etam$,
where $\xim(E, I) > \etam(E, I) >0$ are the two roots of $g(s)=\pm I$;
the intersections of the dashed lines with the solid curve, in Figure~2a,
gives the location of $\xim$ and $\etam$.
The orbits fill a lenticular region, bounded by the 
parabolas $\xi=\xi_m$ and $\eta=-\eta_m$; we call these {\em lens} orbits. 
As the shaded region of
Figure~2b indicates, the lens orbits can visit the origin; {\em they are to cusps with black holes, what box orbits are to analytic, triaxial cores}.
When $|I|=GM$, its maximum value, the shaded region collapses to 
an interval of the $y$--axis, $0\leq y\leq \xim/2$. 
For negative values of $I$, the roles of $\xi$ and $\eta$ are interchanged.

\ni (ii) $E > 0\:$: As seen in Figure~2c, $g(s)$ is no longer a monotonic
function of $s$; a minimum, $-\Im < 0$, is attained at $s=[E/2K(2-\gamma)]^{1/(1-\gamma)}$.
For $|I| < GM$, we again obtain lens orbits. But 
a new type of orbits appears for $GM < |I| \leq \Im$;
the shaded region of Figure~2d shows the region filled by one such
orbit for a typical value of $I$, which is taken to be positive.
These are centrophobic box--like
orbits; we call these {\em boxlets}, borrowing the term from 
Miralda--Escud\'e \& Schwarzschild~(1989).  
When $I$ takes its maximum value equal to $\Im$, the orbit collapses to one section of a constant--$\eta$ curve lying between two halves of the $\xi=\xim$
curve. This is a
$2:1$ resonant {\em banana} orbit, which now emerges
naturally as a section of a parabola!

It is worth noting that loop orbits are completely absent. 
In the absence of the black hole ($M=0$), the lens orbits disappear, 
leaving the boxlets as the only
family of orbits for cusps without black holes. 

\begin{figure}
\epsfxsize=3.3truein\epsfbox[49 167 565 690]{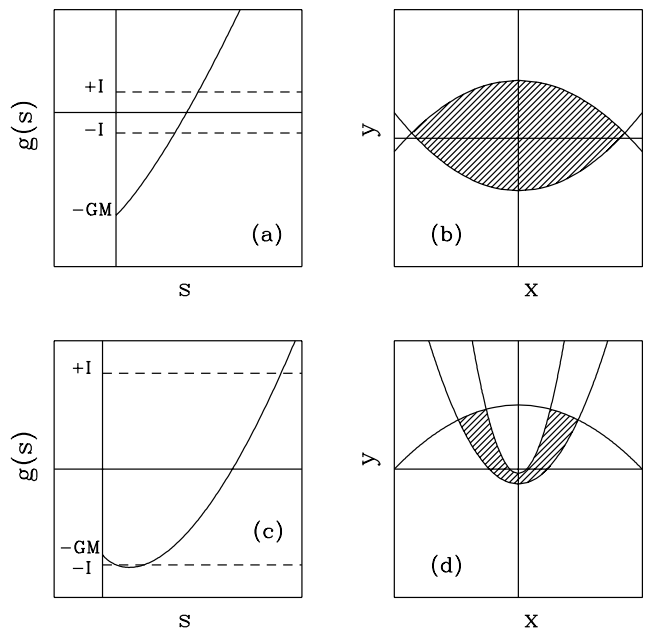}
\caption{Orbital Structure: In all figures $\gamma = 0.5$,
and $I$ has been asumed to be positive.
The solid vertical and horizontal
lines in (a) and (c) correspond to the axes $s=0$ and $g(s)=0$ respectively.
Similar lines in (b) and (d) are the axes $x=0$ and $y=0$ respectively.
(a) $E <0$, The solid curve is
$g(s)$, which intersects the dashed lines $g=\pm I$ at 
$s=\xim$ and $s=\etam$ respectively. (b) {\em Lens} orbits
are restricted to the shaded region of the $x-y$ plane, which is bounded by the curves $\xi=\xim$
and $\eta=-\etam$. 
(c) When $E>0$, the minimum of $g(s)$ is shifted away from 
$s=0$. For $GM < I < \Im$, the dashed line at $-I$ intersects the 
$g(s)$ curve at two non--zero values of $\eta$. This property forces the
{\em boxlets} to be centrophobic, as shown in (d).}
\end{figure}

\section{CONCLUSIONS}
The potential--density pairs presented in this paper are 
the only examples known to us,  of non--axisymmetric 
distributions
of matter that have density cusps  giving
rise to separable motion. The potentials are of St\"ackel form, and 
owe their existence to the geometric properties of the  
parabolic coordinate curves.\footnote{St\"ackel
potentials in elliptic coordinates do not appear to describe density
cusps of interest to dynamics at the centres of elliptical galaxies!}
As a bonus we can also include a central black hole
of arbitrary mass, while maintaining the separable nature.
The discs are significantly non--axisymmetric, with the major axes being
more than twice as long as the minor axes.
The cusps can be steep, with logarithmic slopes of the surface density
lying between $-1$ and $0\,$. 
Different discs may be superposed to obtain mass models that are
not scale--free, yet whose potentials are of St\"ackel form. 
It is a surprising fact that the presence of a black hole
stabilises a family of orbits, the {\em lenses}, which can approach
arbitrarily close to the centre!\footnote{Of course, a star that approaches 
too close to the black hole will be tidally shreded.}
The lenses do not exist in 
the absence of the black hole---the centrophobic {\em boxlets} are the only stable family of
orbits for non--axisymmetric cusps, without a central black hole. 
From the geometry of
the constant $\xi$ and $\eta$ curves, it is evident that the lenses and boxlets
can be elongated in the same sense as the mass distribution, a
property that is favourable to the construction of non axisymmetric, 
self--consistent equilibria.
Although we have limited discussion to discs 
that are symmetric with
respect to reflections about both $x$ and $y$ axes, it is a 
simple matter to break the symmetry with respect to the $x$ axis;
weighting the functions $\scrF_{+}(\xi)$ and $\scrF_{-}(\eta)$ differently in equations~(\ref{ffstackel}) will  produce a family of {\em lopsided}, cuspy
discs. In fact, choosing functional forms for $\scrF_{+}(\xi)$ and $\scrF_{-}(\eta)$ different from the ones given in equations~(\ref{ffstackel})
presents us with a variety of separable  potentials 
(which, however,  may not always be 
derivable from positive densities).
The separable potentials we have presented might also be  
a suitable point of departure for perturbative analyses of dynamics
in more realistic models of non--axisymmetric, lopsided discs.
The scale--free nature simplifies the problem, but we 
clearly need to construct more realistic, finite mass configurations.
Generalization to triaxial cusps, and construction of self--consistent
equilibria remain as some of the outstanding questions.
The restoration of integrability, and the existence of favourably oriented
orbits in the idealized problem we have considered
suggests that it is perhaps premature to conclude that
elliptical galaxies with central density cusps, and black holes 
cannot be triaxial.

\section{ACKNOWLEDGMENTS}

We thank Rajaram Nityananda for stimulating discussions, Scott Tremaine
for useful comments, and the Raman Research Institute and the Indian Railways
for hospitality, while this work was in progress.
JT wishes to thank IUCAA for support and 
hospitality at Pune, where this work began.

\label{lastpage}

\begin{thebibliography}{}

\bibitem{}Bertola,~F., \& Capaccioli,~M., 1975, ApJ, 200, 439

\bibitem{}Binney,~J.J., 1976, MNRAS, 177, 19

\bibitem{}Binney,~J.J., 1978, MNRAS, 183, 501

\bibitem{}Binney,~J.J., \& Tremaine,~S., 1987, Galactic Dynamics,
Princeton University Press, Princeton, Problem~2--13, p.~102

\bibitem{}de~Zeeuw,~P.T., \& Lynden--Bell,~D., 1985, MNRAS, 215, 713

\bibitem{}de~Zeeuw,~P.T., 1985, MNRAS, 216, 273

\bibitem{}Ferrarese,~L., van den Bosch,~F.C., Ford,~H.C., Jaffe,~W.,
\& O'Connell,~R.W., 1994, AJ, 108, 1598

\bibitem{}Gerhard,~O.E., \& Binney,~J.J., 1985, MNRAS, 216, 467

\bibitem{}Illingworth,~G.D.,1977, ApJ, 218, L43

\bibitem{}Kormendy,~J., \& Richstone,~D., 1995, ARA\& A, 33, 581, 

\bibitem{}Kuzmin,~G.G., 1973, ed. T.B.~Omarov, in Dynamics of Galaxies and Clusters
(Alma Ata: Akad. Nauk. Kaz. SSR), p.~71 (transl. in 
IAU Symposium 127, Structure and Dynamics of Elliptical Galaxies,
ed. P.T.~de Zeeuw (Dordrecht: Reidel), p.~553).

\bibitem{}Landau,~L.D., \& Lifshitz,E.M., 1976, Mechanics, 3rd Edition,
Pergamon Press, Oxford

\bibitem{}Lauer,~T.R., Ajhar,~E.A., Byun,~Y.I., Dressler,~A., Faber,~S.M.,
Grillmair,~C., Kormendy,~J., Richstone,~D.O., \& Tremaine,~S., 1995,
AJ, 110, 2622

\bibitem{} Merritt,~D., 1996, Sci, 271, 337

\bibitem{}Miralda--Escud\'e,~J., \& Schwarzschild,~M., 1989, ApJ, 339, 752

\bibitem{}M{\o}ller,~P., Stiavelli,~M., \& Zeilinger,~W.W., 1995,
MNRAS, 276, 979

\bibitem{}Morse,~P.M., \& Feshbach,~H., 1953, Methods of Theoretical
Physics, McGraw Hill, New York, Ch.~5

\bibitem{}Schwarzschild,~M., 1979, ApJ, 232, 236

\bibitem{}Schwarzschild,~M., 1982, ApJ, 263, 599

\bibitem{}Statler,~T.S., 1987, ApJ, 321, 113


\end{thebibliography}
\end{document}